# When We Can Trust Computers (and When We Can't)

## Peter V. Coveney[1,2*] Roger R. Highfield[3]

[1]Centre for Computational Science, University College London, Gordon Street, London WC1H 0AJ, UK,
https://orcid.org/0000-0002-8787-7256
[2]Institute for Informatics, Science Park 904, University of Amsterdam, 1098 XH Amsterdam, Netherlands
[3]Science Museum, Exhibition Road, London SW7 2DD, UK, https://orcid.org/0000-0003-2507-5458

## Abstract

With the relentless rise of computer power, there is a widespread expectation that computers can solve the most pressing problems of science, and even more besides. We explore the limits of computational modelling and conclude that, in the domains of science and engineering that are relatively simple and firmly grounded in theory, these methods are indeed powerful. Even so, the availability of code, data and documentation, along with a range of techniques for validation, verification and uncertainty quantification, are essential for building trust in computer generated findings. When it comes to complex systems in domains of science that are less firmly grounded in theory, notably biology and medicine, to say nothing of the social sciences and humanities, computers can create the illusion of objectivity, not least because the rise of big data and machine learning pose new challenges to reproducibility, while lacking true explanatory power. We also discuss important aspects of the natural world which cannot be solved by digital means. In the long-term, renewed emphasis on analogue methods will be necessary to temper the excessive faith currently placed in digital computation.

## Introduction

Scientists and engineers take reproducibility seriously[1] for reasons that, though obvious, are well worth restating. Research relies on a never-ending dialogue between hypothesis and experiment, a conversation that advances more quickly towards understanding phenomena when not distracted by false leads.

Reproducibility and quantifying uncertainty are vital for the design of buildings, bridges, cars, spacecraft and aircraft, along with weather forecasting and a host of other applications where lives depend on performing calculations correctly. Reproducibility is also a welcome corollary of the open science movement that seeks to make all aspects of research transparent and accessible – whether publications, data, physical samples, algorithms, software and documentation. Many have referred to reproducibility as being a 'cornerstone of science'[2–4].

However, problems of reproducibility arise in all fields of science, from biomedicine, migration and climate to advanced materials, fusion derived energy and high energy physics. There are many reasons why "one-off" observations cannot be reproduced, even with the same methods, owing to the aleatoric, or random, nature of many phenomena. In such instances, reliable statistical measures are essential to convert measurements into robust findings. False correlations are usually washed away when, over time, they are scrutinised by more systematic, bigger and better-designed studies.

*Author for correspondence: p.v.coveney@ucl.ac.uk.

†Centre for Computational Science, University College London,
Gordon Street, London WC1H 0AJ, UK

The extent to which reproducibility is an issue for computer modelling is more profound and convoluted however, depending on the domain of interest, the complexity of the system, the power of available theory, the customs and practices of different scientific communities, and many practical considerations, such as when commercial considerations are challenged by scientific findings.[5]

For research on microscopic and relatively simple systems, such as those found in physics and chemistry, for example, theory – both classical and quantum mechanical - offers a powerful way to curate the design of experiments and weigh up the validity of results. In these and other domains of science that are grounded firmly on theory, computational methods more easily help to confer apparent objectivity, with the obvious exceptions of pathological science[6] and fraud[7]. For the very reason that the underlying theory is established and trusted in these fields, there is perhaps less emphasis than there should be on verification and validation ("solving the equations right" and "solving the right equations", respectively[8]) along with uncertainty quantification—collectively known by the acronym VVUQ. By comparison, in macroscopic systems of interest to engineers, applied mathematicians, computational scientists and technologists and others who have to design devices and systems that actually work, and which must not put people's lives in jeopardy,  VVUQ, is a way of life -  in every sense - to ensure that simulations are credible.

This VVUQ philosophy underpins advances in computer hardware and algorithms that improve our ability to model complex processes using techniques such as finite element analysis, and computational fluid dynamics for end-to-end simulations in virtual prototyping and to create digital twins.[9] There is a virtuous circle in VVUQ, where experimental data hone simulations, while simulations hone experiments and data interpretation. In this way, the ability to simulate an experiment influences validation by experiment.

In other domains, however, notably biology and biomedical sciences, theories have rarely attained the power and generality of physics. The state space of biological systems tends to be so vast that detailed predictions are often elusive, and VVUQ is less well established, though that is now changing rapidly as, for example, models and simulations begin to find clinical use[10].

Despite the often stated importance of reproducibility, researchers still find various ways to unwittingly fool themselves and their peers[11]. Data dredging—also known as blind big data, data fishing, data snooping, and p-hacking—seeks results that can be presented as statistically significant, without any knowledge of the structural characteristics of the problem, let alone first devising a hypothesis about the underlying mechanistic relationships. While corroboration or indirect supporting evidence may be reassuring, when taken too far it can lead to the interpretation of random patterns as evidence of correlations, and to conflation of these correlations with causative effects.

Spurred on by the current reward and recognition systems of academia, it is easier and very tempting to quickly publish one-off findings which appear transformative, rather than invest additional money, energy and time to ensure that these one-off findings are reproducible.  As a consequence, a significant number of 'discoveries' turn out to be unreliable because they are more likely to depend on small populations, weak statistics and flawed analysis[12–16]. There is also a temptation to carry out *post hoc* rationalisation or HARKing, 'Hypothesizing After the Results are Known' and to invest more effort into explaining away unexpected findings than validating expected results. Most contemporary research depends heavily on computers which generate numbers with great facility. Ultimately, though, computers are themselves tools that are designed and used by people.  Because human beings have a capacity for self-deception[17], the datasets and algorithms that they create can be subject to unconscious biases of various kinds, for example in the way data are collected and curated in data dredging activities, a lack of standardized data analysis workflows[18], or the selection of tools that generate promising results, even if their use is not appropriate in the circumstances.

No field of science is immune to these issues, but they are particularly challenging in domains where systems are complex and many dimensional, weakly underpinned by theoretical understanding, and exhibit non-linearity, chaos and long-range correlations. With the rise of digital computing power, approaches predicated on big data, machine learning (ML) and artificial intelligence (AI) are frequently deemed to be indispensable. ML and AI are increasingly used to sift experimental and simulation data for otherwise hidden patterns that such methods may suggest are significant. Reproducibility is particularly important here because these forms of data analysis play a disproportionate role in producing results and supporting conclusions.

Some even maintain that big data analyses can do away with the scientific method.[19] However, as data sets increase in size, the ratio of false to true correlations increases very rapidly, so one must be able to reliably distinguish false from true if one is able to find robust correlations. That is difficult to do without a reliable theory underpinning the data being analysed. We, like others[20], argue that the faith placed in big data analyses is profoundly misguided: to be successful, big data methods must be more firmly grounded on the scientific method.[21] Far from being a threat to the scientific method, the weaknesses of blind big data methods serve as a timely reminder that the scientific method remains the most powerful means we have to understand our world.

**Reproducibility**

In science, unlike politics, it does not matter how many people say or agree about something: if science is to be objective, it has to be reproducible ("within the error bars"). Observations and "scientific facts and results" cannot depend on who is reporting them but must be universal. Consensus is the business of politics and the scientific equivalent only comes after the slow accumulation of unambiguous pieces of empirical evidence (albeit most research and programmes are still funded on the basis of what the majority of people on a review panel thinks is right, so that scientists who have previously been successful are more likely to be awarded grants[22,23].)

There is some debate about the definition of reproducibility[24]. Some argue that replicability is more important than reproducibility. Others maintain that the gold standard of research should be 're-testability', where the result is replicated rather than the experiment itself, though the degree to which the 'same result' can emerge from different setups, software and implementations is open to question.[25]

By reproducibility we mean the repetition of the findings of an experiment or calculation, generally by others, providing independent confirmation and confidence that we understand what was done and how, thus ensuring that reliable ideas are able to propagate through the scientific community and become widely adopted. When it comes to computer modelling, reproducibility means that the original data and code can be analysed by any independent, sceptical investigator to reach the same conclusions. The status of all investigators is supposedly equal and the same results should be obtained regardless of who is performing the study, within well-defined error bars – that is, reproducibility must be framed as a statistically robust criterion because so many factors can change between one set of observations and another, no matter who performs the experiment.

The uncertainties come in two forms: (i) "epistemic", or systematic errors, which might be due to differences in measuring apparatus; and (ii) "aleatoric", caused by random effects. The latter typically arise in chaotic dynamical systems which manifest extreme sensitivity to initial conditions, and/or because of variations in conditions outside of the control of an experimentalist.

By seeking to control uncertainty in terms of a margin of error, reproducibility means that an experiment or observation is robust enough to survive all manner of scientific analysis. Note, of course, that reproducibility is a necessary but not a sufficient condition for an observation to be deemed scientific. In the scientific enterprise, a single result or measurement can never provide definitive resolution for or against a theory. Unlike mathematics, which advances when a proof is published, it takes much more than a single finding to establish a novel scientific insight or idea. Indeed, in the Popperian view of science, there can be no final vindication of the validity of a scientific theory: they are all provisional, and may eventually be falsified. The extreme form of the modern machine-learners' pre-Baconian view stands in stark opposition to this: there is no theory at all, only data, and success is measured by how well one's learning algorithm performs at discerning correlations within these data, even though many of these correlations will turn out to be false, random or meaningless.

Moreover, in recent years, the integrity of the scientific endeavour has been open to question because of issues around reproducibility, notably in the biological sciences. Confidence in the reliability of clinical research has, for example, been under increasing scrutiny.[5] In 2005, John P. A. Ioannidis wrote an influential article about biomedical research, entitled "Why Most Published Research Findings are False", in which he assessed the positive predictive value of the truth of a research finding from values such as threshold of significance and power of the statistical test applied.[26] He found that the more teams were involved in studying a given topic, the less likely the research findings from individual studies turn out to be true. This seemingly paradoxical corollary follows because of the scramble to replicate the most impressive "positive" results and the attraction of refuting claims made in a prestigious journal, so that early replications tend to be biased against the initial findings. This 'Proteus phenomenon' has been observed as an early sequence of extreme, opposite results in retrospective hypothesis-generating molecular genetic research[27], although there is often a fine line to be drawn between contrarianism, wilful misrepresentation and the scepticism ('nullius in verba') that is the hallmark of good science.[28]

Such lack of reproducibility can be troubling. An investigation of 49 medical studies undertaken between 1990–2003 - with more than 1000 citations in total - found that 16% were contradicted by subsequent studies, 16% found stronger effects than subsequent studies, 44% were replicated, and 24% remained largely unchallenged.[29] In psychological science, a large portion of independent experimental replications did not reproduce evidence supporting the original results despite using high-powered designs and original materials when available.[30] Even worse performance is found in cognitive neuroscience.[13]

Scientists more widely are routinely confronted with issues of reproducibility: a May 2016 survey in *Nature* of more than 1576 scientists reported that more than 70% had tried and failed to reproduce another scientist's experiments, and more than half had failed to reproduce their own experiments.[31] This lack of reproducibility can be devastating for the credibility of a field.

**Modelling and Simulation**

Computers are critical in all fields of data analysis and computer simulations need to be reliable - validated, verified, and their uncertainty quantified - so that they can feed into real world applications and decisions be they governmental policies dealing with pandemics, for the global climate emergency, the provision of food and shelter for refugee populations fleeing conflicts, creation of new materials, the design of the first commercial fusion reactor, or to assist doctors to test medication on a virtual patient before a real one.

Reproducibility in computer simulations would seem trivial to the uninitiated: enter the same data into the same program on the same architecture and you should get the same results. In practice, however, there are many barriers to overcome to ensure the fidelity of a model in a computational environment.[32] Overall, it can be challenging if not impossible to test the claims and arguments made by authors in published work without access to the original code and data, and even in some instances the machines the software ran on. One study of what the authors dubbed 'weak repeatability' examined 402 papers with results backed by code and found that, for one third, they were able to obtain the code and build it within half an hour, while for just under half they succeeded with significant extra effort. For the remainder, it was not possible to verify the published findings. The authors reported that some researchers are reluctant to share their source code, for instance for commercial and licensing reasons, or because of dependencies on other software, whether due to external libraries or compilers, or because the version they used in their paper had been superseded, or had been lost due to lack of backup. Many detailed choices in the design and implementation of a simulation never make it into published papers. Frequently, the principal code developer has moved on, the code turns out to depend on exotic hardware, there is inadequate documentation, and/or the code developers say that they are too busy to help.[33] There are some high-profiles examples of these issues, from disclosure of climate codes and data,[34] to delays in sharing codes for COVID-19 pandemic modelling.[35] If the public are to have confidence in computing models that could directly affect them, transparency, openness and the timely release of code and data are critical.

In response to this challenge, there have been various proposals to allow scientists to openly share code and data that underlie their research publications: RunMyCode [runmycode.org] and, perhaps better known, GitHub [github.com]; SHARE, a web portal to create, share, and access remote virtual machines that can be cited from research papers to make an article fully reproducible and interactive;[36] PaperMâché, another means to view and interact with a paper using virtual machines;[37] various means to create 'executable papers'[38,39]; and a verifiable result identifier (VRI), which consists of trusted and automatically generated strings that point to publicly available results originally created by the computational process.[40]

In addition to external verification, there are many initiatives to incorporate verification and validation into computer model development, along with uncertainty quantification techniques to verify and validate the models.[41] In the United States, for example, the American Society of Mechanical Engineers has a Standards Committee for the development of verification and validation V&V procedures for computational solid mechanics models,[42] guidelines and recommended practices have been developed by the National Aeronautics and Space Administration (NASA);[43] the US Defense Nuclear Facilities Safety Board backs model V&V for all safety-related nuclear facility design, analyses, and operations, while various groups within the DOE laboratories (including Sandia, Los Alamos, and Lawrence Livermore) are conducting research in this area.[44] In Europe, the VECMA (Verified Exascale Computing for Multiscale Applications) project[45] is developing software tools that can be applied to many research domains, from the laptop to the emerging generation of exascale supercomputers, in order to validate, verify, and quantify the uncertainty within highly diverse applications.

The major challenge faced by the state of the art is that many scientific models are multiphysics in nature, combining two or more kinds of physics, for instance to simulate the behaviour of plasmas in tokamak nuclear fusion reactors[46], electromechanical systems[47] or in food processing[48]. Even more common, and more challenging, many models are also multiscale, which require the successful convergence of various theories that operate at different temporal and/or spatial scales. They are widespread at the interface between various fields, notably physics, chemistry and biology. The ability to integrate macroscopic universality and molecular individualism is perhaps the greatest challenge of multiscale modelling[49]. As one example, we certainly need multiscale models if we are to predict the biology, medicine that underpin the behaviour of an individual person. Digital medicine is increasingly important and, as a corollary of this, there have been calls for steps to avoid a reproducibility "crisis" of the kind that has engulfed other areas of biomedicine.[50]

Although there are many kinds of multiscale modelling, there now exist protocols to enable the verification, validation, and uncertainty quantification of multiscale models.[51] The VECMA toolkit[52], which is not only open source but whose development is also performed openly, has many components: FabSim3, to organise and perform complex remote tasks; EasyVVUQ, a Python library designed to facilitate verification, validation and uncertainty quantification for a variety of simulations[53,54]; QCG Pilot Job, to provide the efficient and reliable execution of large number of computational jobs; QCG-Now, to prepare and run computational jobs on high performance computing machines;  QCG-Client, to provide support for a variety of computing jobs, from simple ones to complex distributed workflows; EasyVVUQ-QCGPilotJob, for efficient, parallel execution of demanding EasyVVUQ scenarios on high performance machines; and MUSCLE 3, to make creating coupled multiscale simulations easier, and to then enable efficient uncertainty quantification of such models.

The VECMA toolkit is already being applied in several circumstances: climate modelling, where multiscale simulations of the atmosphere and oceans are required; forecasting refugee movements away from conflicts, or as a result of climate change, to help prioritise resources and investigate the effects of border closures and other policy decisions[55]; for exploring the mechanical properties of a simulated material at several length and time scales with verified multiscale simulations; and multiscale simulations to understand the mechanisms of heat and particle transport in fusion devices, which is important because the transport plays a key role in determining the size, shape and more detailed design and operating conditions of a future fusion power reactor, and hence the possibility of extracting almost limitless energy; and verified simulations to aid in the decision-making of drug prescriptions, simulating how drugs interact with a virtual version of a patient's proteins,[56] or how stents will behave when placed in virtual versions of arteries.[57]

**Big Data, Machine Learning and Reproducibility**

Recent years have seen an explosive growth in digital data accompanied by the rising public awareness that their lives depend on "algorithms", though it is plain to all that any computer code is based on an algorithm, without which it will not run. Under the banner of artificial intelligence and machine learning, many of these algorithms seek patterns in those data. Some – emphatically not the authors of this paper - even claim that this approach will be faster and more revealing than modelling the underlying behaviour notably by the use of conventional theory, modelling and simulation.[58] This approach is particularly attractive in disciplines traditionally not deemed suitable for mathematical treatment because they are so complex, notably life and social sciences, along with the humanities.

However, to build a machine-learning system, you have to decide what data you are going to choose to populate it. That choice is frequently made without any attempt to first try to understand the structural characteristics that underlie the system of interest, with the result that the "AI system" produced strongly reflects the limitations or biases (be they implicit or explicit) of its creators.

Moreover, there are four fundamental issues with big data that are frequently not recognised by practitioners[58]: complex systems are strongly correlated, so they do not generally obey Gaussian statistics; no datasets are large enough for systems with strong sensitivity to rounding or inaccuracies; correlation does not imply causality; and too much data can be as bad as no data: although computers can be trained on larger datasets than the human brain can absorb, there are fundamental limitations to the power of such datasets (as one very real example, mapping genotype to phenotype is far from straightforward), not least due to their digital character.

All machine-learning algorithms are initialised using (pseudo) random number generators and have to be run vast numbers of times to ensure that their statistical predictions are robust. However, they typically make plenty of other assumptions, such as smoothness (i.e. continuity) between data points. The problem is that nonlinear systems are often anything but smooth, and there can be jumps, discontinuities and singularities.

Not only the smoothness of behaviour but also the forms of distribution of data regularly assumed by machine learners are frequently unknown or untrue in complex systems. Indeed, many such approaches are distribution free, in the sense that there is no knowledge provided about the way the data being used is distributed in a statistical sense.[58] Often, a Gaussian ("normal") distribution is assumed by default; while this distribution plays an undeniable role across all walks of science it is far from universal. Indeed, it fails to describe most phenomena where complexity holds sway because, rather than resting on randomness, these typically have feedback loops, interactions and correlations.

Machine learning is often used to seek correlations in data. But in a real-world system, for instance in a living cell that is a cauldron of activity of 42 million protein molecules[59], can we be confident that we have captured the right data? Random data dredging for complex problems is doomed to fail where one has no idea which variables are important. In these cases, data dredging will always be defeated by the curse of dimensionality – there will simply be far too much data needed to fill in the hyperdimensional space for blind machine learning to produce correlations to any degree of confidence. On top of that, as mentioned earlier, the ratio of false to true correlations soars with the size of the dataset, so that too much data can be worse than no data at all.

There are practical considerations too. Machine-learning systems can never be better than the data they are trained on, which can contain biases 'whether morally neutral as toward insects or flowers, problematic as toward race or gender, or even simply veridical, reflecting the *status quo* distribution of gender with respect to careers or first names'.[60] In healthcare systems, for example, where commercial prediction algorithms are used to identify and help patients with complex health needs, significant racial bias has been found.[61] Cathy O'Niel's book, *Weapons of Math Destruction* is replete with examples of this kind, covering virtually all walks of life, and their harmful impact on modern societies.[62]

Machine learning systems are black boxes, even to the researchers that build them, making it hard for their creators, let alone others, to assess the results produced by these glorified curve-fitting systems. Precise replication would be nearly impossible given the natural randomness in neural networks and variations in hardware and code. That is one reason why blind machine learning is unlikely to ever be accepted by regulatory authorities in medical practice as a basis for offering drugs to patients. To comply with the regulatory authorities such as the US Food and Drug Administration and the European Medicines Agency, the predictions of a ML algorithm are not enough and it is essential that an underlying mechanistic explanation is also provided, one which can explain not only when a drug works but also when it fails, and/or produces side effects.

There are even deeper problems of principle in seeking to produce reliable predictions about the behaviour of complex systems of the sort one encounters frequently in the most pressing problems of twenty-first century science. We are thinking particularly, in life sciences, medicine, healthcare and environmental sciences, where systems typically involve large numbers of variables and many parameters. The question is how to select these variables and parameters to best fit the data. Despite the constant refrain that we live in the age of "Big Data", the data we have available is never enough to model problems of this degree of complexity. Unlike more traditional reductionist models, where one may reasonably assume one has sufficient data to estimate a small number of parameters, such as a drug interacting with a nerve cell receptor, this ceases to be the case in complex and emergent systems, such as modelling a nerve cell itself. The favourite approach of the moment is of course to select machine learning, which involves adjustments of large numbers of parameters inside the neural network "models" used; these can be tuned to fit the data available but have little to no predictability beyond the range of the data used because they do not take into account the structural characteristics of the phenomenon under study. This is a form of overfitting.[63] As a result of the uncertainty in all these parameters, the model itself becomes uncertain as testing it involves an assessment of probability distributions over the parameters and, with nowhere near adequate data available, it is not clear if it can be validated in a meaningful manner.[64] For some related issues of a more speculative and philosophical nature in the study of complexity, see Succi (2019).

Compounding all this, there is a fundamental problem that undermines our faith in simulations which arises from the digital nature of modern computers, whether classical or quantum. Digital computers make use of four billion rational numbers that range from plus to minus infinity, the so-called 'single-precision IEEE floating-point numbers', which refers to a technical standard for floating-point arithmetic established by the Institute of Electrical and Electronics Engineers in the 1950s; they also frequently use double precision floating-point numbers, while half-precision has become commonplace of late in the running of machine learning algorithms.

However, digital computers only use a very small subset of the rational numbers – so-called dyadic numbers, whose denominators are powers of 2 because of the binary system underlying all digital computers – and the way these numbers are distributed is highly nonuniform. Moreover, there are infinitely more irrational than rational numbers, which are ignored by all digital computers because to store any one of them, typically, one would require an infinite memory. Manifestly, the IEEE floating point numbers are a poor representation even of the rational numbers. Recent work by one of us (PVC) in collaboration with Bruce Boghosian and Hongyan Wang at Tufts University, demonstrates that there are major errors in the computer-based prediction of the behaviour of arguably the simplest of chaotic dynamical systems, the generalised Bernoulli map, regardless of the degree of precision of the floating-point numbers used.[65]

Leaving aside the mistaken belief held by some that a very few repeats of, say, a molecular dynamics simulation is any replacement for (quasi) Monte Carlo methods based on ensembles of replicas, these findings strongly suggest that the digital simulation of all chaotic systems, found in models used to predict weather, climate, molecular dynamics, chemical reactions, fusion energy and much more, contain sizeable errors of a nature that hitherto have been unknown to most scientists. By the same token, the use of data from these chaotic simulations to train machine learning algorithms will in turn produce artefacts, making them unreliable.

This shortcoming produced generic errors of up to 20 per cent in the case of the Bernoulli map along with pure nonsense on rare occasions. One might ask why, if the consequences can be so substantial, these errors have not been noticed. The difficulty is that for real world simulations in turbulence and molecular dynamics, for example, there are no exact, closed form mathematical solutions for comparison so the numerical solutions that roll off the computer are simply assumed to be correct. Given the approximations involved in such models, not to speak of the various sources of measurement errors, it is never possible to obtain exact agreement with experimental results. In short, the use of floating point numbers instead of real numbers contributes additional systematic errors in numerical schemes that have not so far been assessed at all[66].

**The Solution**

For modelling, we need to tackle both epistemic and aleatoric sources of error. To deal with these challenges, a number of countermeasures have been put forward: documenting detailed methodological and statistical plans of an experiment ahead of data collection (preregistration): demanding that studies are thoroughly replicated before they are published;[67,68] insisting on collaborations to double check findings;[69] explicit consideration of alternative hypotheses, even processing all reasonable scenarios[70]; the sharing of methods, data, computer code and results in central repositories, such as the Open Science Framework,[71] a free, open platform to support research, enable collaboration and 'team science'[72]; and blind data analysis, where data are shifted by an amount known only to the computer, leaving researchers with no idea what their findings implied until everyone agrees on the analyses and the blindfold is lifted. The role of universities, as in the Brazilian Reproducibility Initiative,[73] is important, along with conferences, such as the World Conferences on Research Integrity [https://wcrif.org/], and the actions of funding agencies, such as the US National Institutes of Health,[74] the UK research councils and the Wellcome Trust,[75] along with the French National Center for Scientific Research (CNRS), which has launched CASCAD, Certification Agency for Scientific Code and Data, [www.cascad.tech], the first public laboratory specialized in the certification of the reproducibility of scientific research.

Machine learning requires special consideration. A survey of 400 algorithms presented in papers at two top AI conferences (the 2013 and 2016 International Joint Conferences on Artificial Intelligence, IJCAI, and the 2014 and 2016 Association for the Advance of Artificial Intelligence, AAAI, conferences[76]) found that only 6% of the presenters shared the algorithm's code.[77] The most commonly-used machine learning platforms provided by big tech companies have poor support for reproducibility.[78] Studies have shown that even if the results of a deep learning model could be reproduced, a slightly different experiment would not support the findings—yet another example of overfitting—which is common in machine learning research. In other words, unreproducible findings can be built upon supposedly reproducible methods.[79]

Rather than continuing to simply fund, pursue and promote 'blind' big data projects, more resources should be allocated to the elucidation of the multiphysics, multiscale and stochastic processes controlling the behaviour of complex systems, such as those in biology, medicine, healthcare and environmental science.[21] Finding robust predictive mechanistic models that provide explanatory insights will be of particular value for machine learning when dealing with sparse and incomplete sets of data, ill-posed problems, exploring vast design spaces to seek correlations and then, most importantly, for identifying correlations. Where machine learning provides a correlation, multiscale modelling can test if this correlation is causal.

There are also demands in some fields for a reproducibility checklist,[80] to make AI reproducibility more practical, reliable and effective. Another suggestion is the use of so-called "Model Cards" – documentation that accompanies trained machine learning models which outline the application domains, the context in which they are being used and their carefully benchmarked evaluation in a variety of conditions, such as across different cultural, demographic, and phenotypic groups;[81] and proposals for best practice in reporting experimental results which permit for robust comparison.[82]

Despite the caveat that computers are made and used by people, there is also considerable interest in their use to design and run experiments, for instance using Bayesian optimization methods, such as in the field of cognitive neuroscience[83] and to model infectious diseases and immunology quantitatively.[84]

When it comes to the limitations of digital computing, research is under way by Boghosian and PVC to find alternative approaches that might render such problems computable on digital computers. Among possible solutions, one that seems guaranteed to succeed is analogue computing, an older idea, able to handle the numerical continuum of reality in a way that digital computers can only approximate.[85]

**Synthesis & Conclusion**

In the short term, notably in the biosciences, better data collection, curation, validation, verification and uncertainty quantification procedures of the kind described here, will make computer simulations more reproducible, while machine learning will benefit from a more rigorous and transparent approach. The field of big data and machine learning has become extremely influential but without big theory it remains dogged by a lack of firm theoretical underpinning ensuring its results are reliable.[21] Indeed, we have argued that in the modern era in which we aspire to describe really complex systems, involving many variables and vast numbers of parameters, there is not sufficient data to apply these methods reliably. Our models are likely to remain uncertain in many respects, as it is so difficult to validate them.

In the medium term, AI methods may, if carefully produced, improve the design, objectivity and analysis of experiments. However, this will always require the participation of people to devise the underlying hypotheses and, as a result, it is important to ensure that they fully grasp the assumptions on which these algorithms are based and are also open about these assumptions.

It is already becoming increasingly clear that 'artificial intelligence' is a digital approximation to reality. Moreover, in the long term, when we are firmly in the era of routine exascale and perhaps eventually also quantum computation, we will have to grapple with a more fundamental issue. Even though there are those who believe the complexity of the universe can be understood in terms of simple programs rather than by means of concise mathematical equations,[86,87] digital computers are limited in the extent to which they can capture the richness of the real world.[85,88] Freeman Dyson, for example, speculated that for this reason the downloading of a human consciousness into a digital computer would involve 'a certain loss of our finer feelings and qualities'[89]. In the quantum and exascale computing eras, we will need renewed emphasis on the analogue world and analogue computational methods if we are to trust our computers.[85]

# Additional Information

**Authors' Contributions**
All authors contributed to the concept and writing of the article.

**Competing Interests**
The authors have no competing interests.

**Funding Statement**
P.V.C. is grateful for funding from the UK EPSRC for the UKCOMES UK High-End Computing Consortium (EP/R029598/1), from MRC for a Medical Bioinformatics grant (MR/L016311/1), the European Commission for the CompBioMed, CompBioMed2 and VECMA grants (numbers 675451, 823712 and 800925 respectively) and special funding from the UCL Provost.

**Acknowledgments**
The authors are grateful for many stimulating conversations with Bruce Boghosian, Daan Crommelin, Ed Dougherty, Derek Groen, Alfons Hoekstra, Robin Richardson & David Wright.